# Imminent earthquake forecasting on the basis of Japan INTERMAGNET stations, NEIC, NOAA and Tide code data analysis

Strachimir Cht. Mavrodiev

Institute for Nuclear Reasurch and Nuclear Energy, BAS, Sofia, Bulgaria

## Abstract

This research presentsone possible way for imminent prediction of earthquakes' magnitude, depth and epicenter coordinates by solving the inverse problem using a data acquisition network system for monitoring, archiving and complex analysis of geophysical variables - precursors. Among many possible precursors the most reliable are the geoelectromagnetic field,the boreholes water level, radon earth-surface concentration, the local heat flow, ionosphere variables, low frequency atmosphere and Earth core waves.The title demonstrates that only geomagnetic data are used in this study.

Within the framework of geomagnetic quake approach it is possible to perform an imminent regional seismic activity forecasting on the basis of simple analysis of geomagnetic data which useanew variable $S_{chtm}$ with dimension surface density of energy. Such analysis of Japan Memambetsu, Kakioka, Kanoya INTERMAGNET stations and NEIC earthquakes data, the hypothesis that the "predicted" earthquake is this with bigest value of the variable $S_{chtm}$ **permit to formulate** an inverse problem (overdetermined algebraic system) for precursor's signals like a function of earthquake's magnitude, depth and distance from a monitoring point

Thus, in the case ofdata acquisition network system existence, which includes monitoring of more than one reliable precursor variables in at least four points distributed within the area with a radius of up to 700 km, there will be enough algebraic equations for calculation ofimpending earthquake's magnitude, depth and distance, solving the overdetermined algebraic system.

## 1. Introduction

It is well known now that the "when, where and how" earthquake's prediction problem cannot be solved by analyzing only the earthquakes data base [1]-[5].

The role of the Sun- Moon Earth tides as possible earthquake's triggerhas been analyzed in [6] - [13]. However the conclusion that the earthquake's time is correlated with the time of tidal extremesis not exact, because in some cases the beginning and the extremes of earthquakes do not coincide. There is an extreme but not an earthquake.

The role of the atmospheric and ionosphere electromagnetic phenomena which can serve as earthquake's precursors in the last time has been researched in many studies.

Physical models of the phenomena observed were proposed in[14], reliability of predictions were analyzed in [15, 16].

The heat release as earthquake's precursor was researched in [17].

The variations of regional water-table reflect fast deformational cycles in lithosphere and may also serve as an earthquake's precursor as one was demonstrated by G.S. Vartanyan [18]. The comparison of the daily geomagnetic fluctuations (geomagnetic quakes) and underground water level demonstrates that borehole water level data may serve as an imminent regional earthquake's precursor in the Caucasus region [19].

The analysis of data for radon concentrations and its fluctuations in the atmosphere and ground-water has been demonstrated in many studies - see for example [20], [21]. The most accepted result is that anomalous (increased regional concentration) of the radon emission can serve as a precursor of an earthquake.

The research of the correlation between variations of geo-electromagnetic field and impending earthquakes has a long-time history-[22] -[36].

A comparative analysis of the two measured values in time of geomagnetic field with the calculation of the standard deviation (dispersion) in the same subintervals - periods of time allowed offering geomagnetic quake as an earthquake precursor [36].

The calculation of the differences (*DayDiff*) between the times of the earthquakes occurred in the region around the monitoring point and the nearest time of tide extremes permit to build the distribution of *DayDiff*. It was established that this distribution is described well by Gauss curve with a certain width $W_{all}$.

Introducing a new variable $S_{ChtM}$ with dimension surface energy density, which is a function of earthquake's magnitude, depth and distance to the monitoring point

$$S_{ChtM}(Mag, Depth, Distance)$$

and the calculation of its value in the monitoring point permits to classify the earthquakes occurred in the monitoring region and in the time period around tide extremes time.

The distribution of *DayDiff* for earthquakes with the biggest *value of* $S_{ChtM}$ is also described with Gauss curve, but with less width $W_{pr}$.

In the paper [37] the *DayDiff* for all world's 628873 earthquakes, occurred in the period 1981- 2013, with *Mag*>= 3 ((International Seismological Centre, http://www.isc.ac.uk/data )was calculated and the distribution, described by Gauss curve with width $W_{all}$=4.46+/-0.22.

The distributions of *DayDiff* for earthquakes with the biggest $S_{chtm}$ calculate from the data of INTERMAGNET stations PAG (Panagurichte, BAS, Bulgaria- Jan1, 2008- Jan19, 2013), SUA (SUA, Romania, Jan1, 2008- Jan17, 2013) and AQU ( L'Aquila, Italy, Jan1, 2008- May 30, 2013) were described by Gauss curves with widths 4.22+/-0.62, 4.11+/-0.51 and 4.28+/-0.67. So, one can say that the appearance of geomagnetic quake forecasts that in the next period around time of tide extreme and monitoring point region means an increase in the seismic activity.

There is a simple intuitive physical explanation [36],[45]of the fact that a geomagnetic quake is an earthquake's precursor:
- The increase of the strain before an earthquake is accompanied by electrochemical and electro kinetic effects which generate Earth electrical currents in the epifocal volume;

- These currents, which can be identified using the geomagnetic quake approach.
  The earthquake's preparing continues as follow:
- The preliminary stage of an earthquake is accompanied by negative divergence of the energy due to increased dissipation of elastic tidal waves;
- The maximum of two time daily Tide's acceleration lead to the transformation of this non- equilibrium state to a new balance that is closer to bifurcation, which explains the role of Tides as an earthquake's trigger.

There is the hope that including the above described research of regional earthquake's precursors in the common approach for solving the earthquake prediction problem (see the paper [38] and references there) will lead us to a solution.

In section 2 is describing the approach for forecasting of imminent regional seismic activity on the basis of Japan geomagnetic data and Sun- Moon Earth tide tode data. In section 3 is demonstrated the reliability of geomagnetic quake approach for the regions (700 km) of Memambetsu, Kakioka, Kanoya stations. In section 4 is presented the description of precursor signal as a function of earthquake's magnitude, depth and distance. In section 4 is presented the formulation of inverse problem for forecasting the magnitude, depth and epicenter coordinates of regional imminent earthquake

. In Aplication 1, Table 2 are present data for the stations, earthquake's date, latidude, longitude, depth, magnitude, the value of $S_{ChtM}$ [J/km$^2$], the distance from station [km], the difference between the predicted time and the time of occurred earthquake [day], the values ofexperimetal and model precursor signal and it difference (Expt – Th). In Aplication 2 is presented the FORTRAN version of precursor signal function ***PrecSigTh(Mag,Depth,Distance)***

## 2. Forecasting of Imminent Japan Regional Seismic Activity on the Basis of Geomagnetic and Sun- Moon Earth Tide Code Data

In this paragraph the data acquisition system for archiving, visualization and analysis [41], [42] in a case of Japan geomagnetic data is presented [41], [42].

### 2.1. Description of the Approach- Figure 1.

The data used:
- the Japan INTERMAGNET geomagnetic stations MMB (Memambetsu, Lat 43.907$^o$ N, Lon 144.193$^o$ E, Altitude = 42 m), KAK (Kakioka, Lat 36.232$^o$ N, Lon 140.186$^o$ E, Altitude = 36 m) KNY (Kanoya, Lat 31.42°N, Lon 130.88$^o$E, Altitude = 107 m) minute data (**http://www.intermagnet.org/**),
- the software for calculation of the daily and minute Earth tide behaviour [39] (Dennis Milbert, NASA, **http://home.comcast.net/~dmilbert/softs/solid.htm**),
- the Earth tide extremes (daily average maximum, minimum and inflexed point) as a trigger of earthquakes,
- the data for World A-indices (**http:/www.swpc.noaa.gov/alerts/a-index.html**).

The geomagnetic signal is calculated as a simple function of relative standard deviations of the components of the geomagnetic vector. The precursor signal is the difference between today and yesterday's geomagnetic signal corrected by the A- indices values. As the increase of precursor signal means increase of geomagnetic field variability, we call such positive leap a geomagnetic quake in analogy with an earthquake. The analysis of the correlation between the earthquakes occurred and the time of Sun- Moon Earth tide extremes on the basis of the variable earthquake's surface energy density $S_{ChtM}$ permits to forecast the imminent regional seismic activity. The calculation of the day differences (*DayDiff*) between the time of the earthquakes occurred and the time of the nearest Tide extreme permits to build the curve of *DayDiff* and its Gauss fit. The comparison of Gauss widths for all the earthquakes occurred and those with the biggest $S_{ChtM}$ is basis for formulation the hypothesis for "predictable" earthquakes.

## 2.2. The Simple Mathematics and Description of Variables

. The simple mathematics for the calculation of the precursor signal, the software for illustrating the reliability of forecasting and its statistic estimation and the variables in Fig. 1 are described as follows.

The Geomagnetic field components $North_m, East_m, Down_m$, $m$=1440, are the minute averaged values of the geomagnetic vector $F$, and the variables $SdNorth_h$, $SdEast_h, dSDown_h$ are their standard deviation, calculated for 1 hour, h=1,..,24):

$$SdNorth_h = \sqrt{\frac{\sum_{m=1}^{60}(\overline{North_h} - North_m)^2}{60}} \qquad (1),$$

where

$$\overline{North_h} = \frac{\sum_{m=1}^{60} North_m}{60} \qquad (2);$$

The geomagnetic signal $GeomHourSig_h$ is the geometrical sum of hour standard deviation normed by the module of hour geomagnetic vector:

$$GeomHourSig_h = \sqrt{\frac{SdNorth_h^2 + SdEast_h^2 + SdDown_h^2}{\overline{North_h}^2 + \overline{East_h}^2 + \overline{Down_h}^2}} \qquad (3);$$

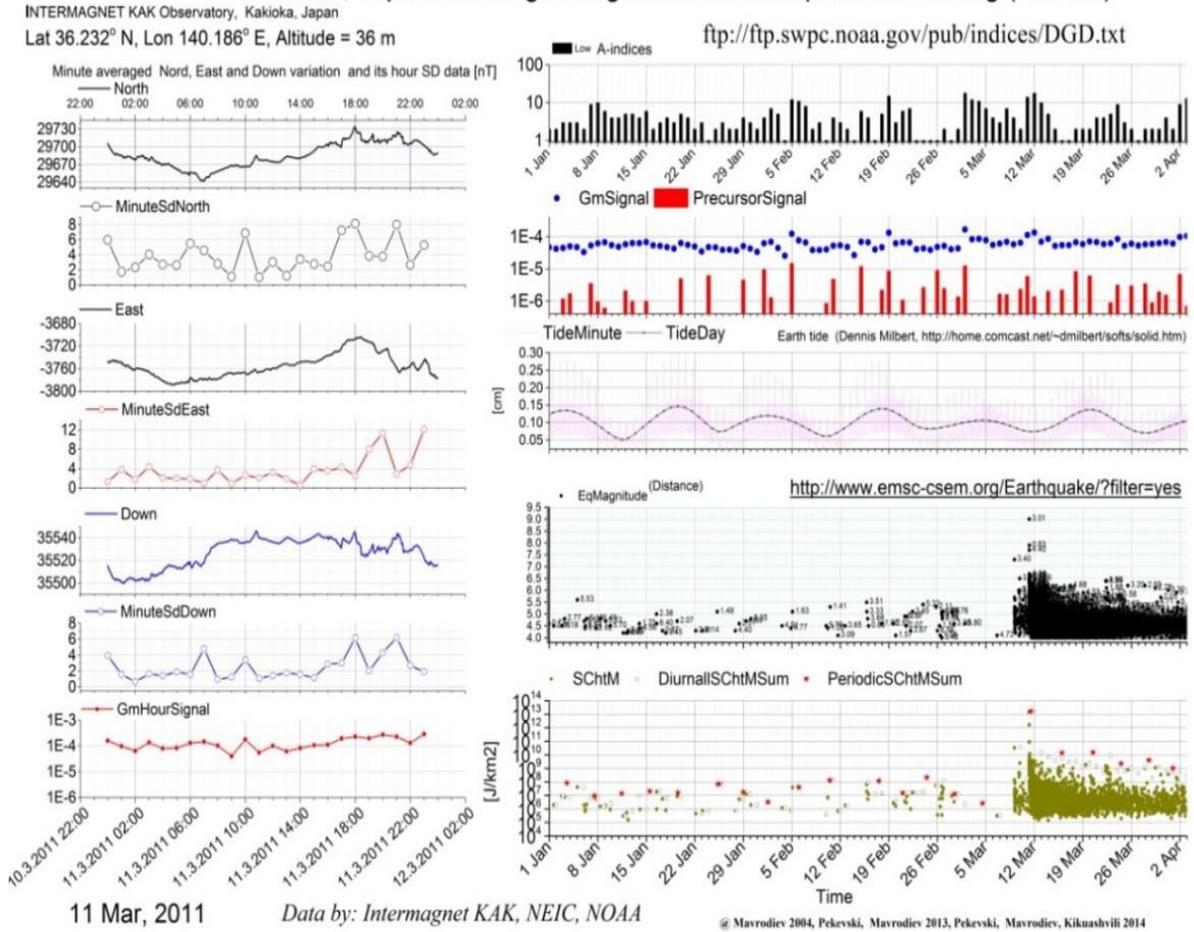

Fig.1. Kakioka diurnal geomagnetic and earthquakes monitoring in the time period around the Fukushima earthquake with geomagnetic field on March 11, 2011.

The $A$ indices are the Low, Medium and High indices, calculated by the NOAA, Space weather prediction center: http://www.swpc.noaa.gov/alerts/a-index.html. In this paper we use $A_{low}$;

The variable $GmSig_{day}$ is the diurnal mean value of $GmHourSig_h$:

$$GeomSig_{day} = \frac{\sum_{h=1}^{24} GeomHourSig_h}{24} \qquad (4)$$

and $PrecursorSig_{day}$

$$PrecursorSig_{day} = 2\frac{GeomSig_{day} - GeomSig_{yesterday}}{Alow_{day} + Alow_{yesterday}} \qquad (5);$$

The indices of earthquake's magnitude value are the distance in hundred km between the epicenter and the monitoring point;

The variable $S_{ChtM}$ is the modified earthquake's surface energy flow density in the monitoring point:

$$S_{ChtM} = \frac{10^{(1.4Mag+4.8)}}{(40+Depth+Distance)^2} \quad [\text{J/km}^2] \quad (6);$$

The variable $PeriodicS_{ChtM}Sum$ [J/km²] is the sum of the variable $S_{ChtM}$ for all earthquakes occurred in the time period +/- 2.7 days before and after the tide extreme in the 700 km region around the monitoring point. Obviously, its value can serve as **estimation of the regional seismic activities for the time period around the tide's extreme;**

The variable $DiurnalS_{ChtM}Sum$ [J/km² per day] is the sum of the variable $S_{ChtM}$, calculated for all earthquakes occurred during the day in the 700 km region around the monitoring point. This variable can serve as a **quantitative measure of diurnal regional seismicity;**

One has to note that the explicit form of the variable $S_{ChtM}$ was established in the framework of inverse problem [41], [46] with the condition to have a clearer correlation between the variable $PrecursorSig_{day}$ and $PeriodicS_{ChtM}Sum$.

The variable $TideMinute$ [cm] is the module of tide vector calculated every 15 minutes;

The variable $TideDay$ [cm] is the diurnal mean value in time calculated in the analogy of *mass center* formulae in many bodies' classical mechanics:

$$Time_{TideDay} = \frac{\sum_{m=1}^{360} mTideDay_m}{\sum_{m=1}^{360} TideDay_m} \quad (7).$$

Note: For seconds and more samples per second, the generalization has to calculate geomagnetic field characteristics for every minute and correspondingly the values of GmSig$_{day}$ have to be the average for 1440 minutes.

The positive value of the variable $PrecursorSig_{day}$ means that the geomagnetic field variability, which is calculated via standard deviations of geomagnetic fileld components, is increasing. **In analogy with earthquake we call such increase a** *geomagnetic quake*.

**As one can see from Fig.1.,after the appearance of a geomagnetic quake, in nine of twelve cases (75%), the regional seismic activity is increasing (the bigger value of the** $PeriodicS_{ChtM}Sum$ **variable) in the time period aroundthe followingtide extreme. So, the geomagneticquake approach described can serve as a forecast of imminent regional seismic activity.**

In Fig.1 the values of the variable *PeriodicS$_{ChtM}$Sum* are calculated not only in the time periods around the extremes, but also in the time period between them. We can see that its values in almost every extreme period are higher.

The use of the above described analysis for a longer time period with calculation of distribution of day difference between the "predicted" earthquakes (earthquakes with the highest value *S$_{ChtM}$*) can demonstrate the reliability of the approach for forecasting imminent regional seismic activity for regions with seismic risk.

## 2. Reliability of Geomagnetic Quake Approachbased on the Analyses of INTERMAGNET Data from MMB (Memambetsu), KAK (Kakioka) and KNY (Kanoya) Stations Located in Japan

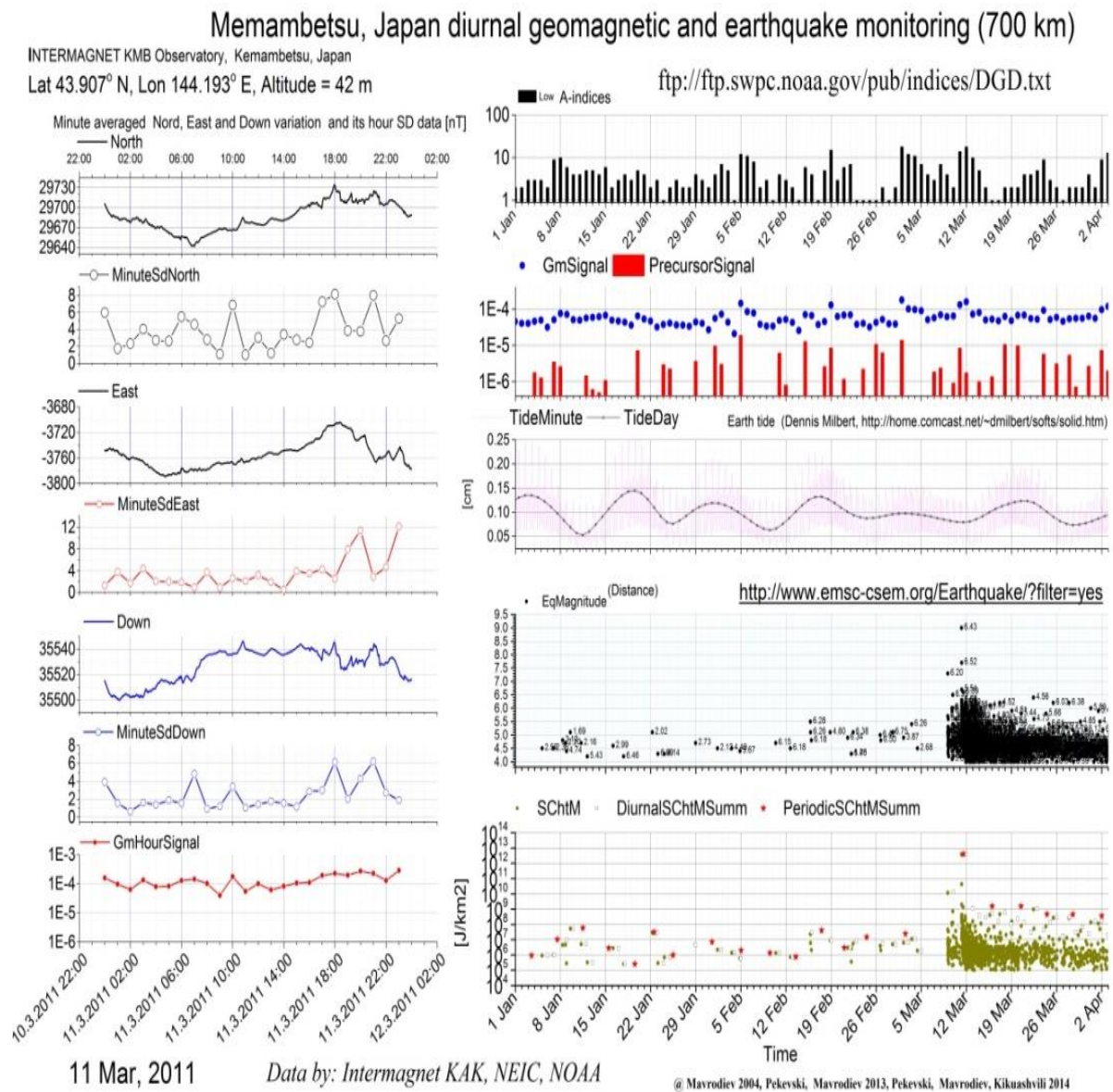

Fig.2. The Memambetsu diurnal geomagnetic and earthquakes monitoringin the period around the time of the Fukushima earthquake with geomagnetic field on March 11, 2011.

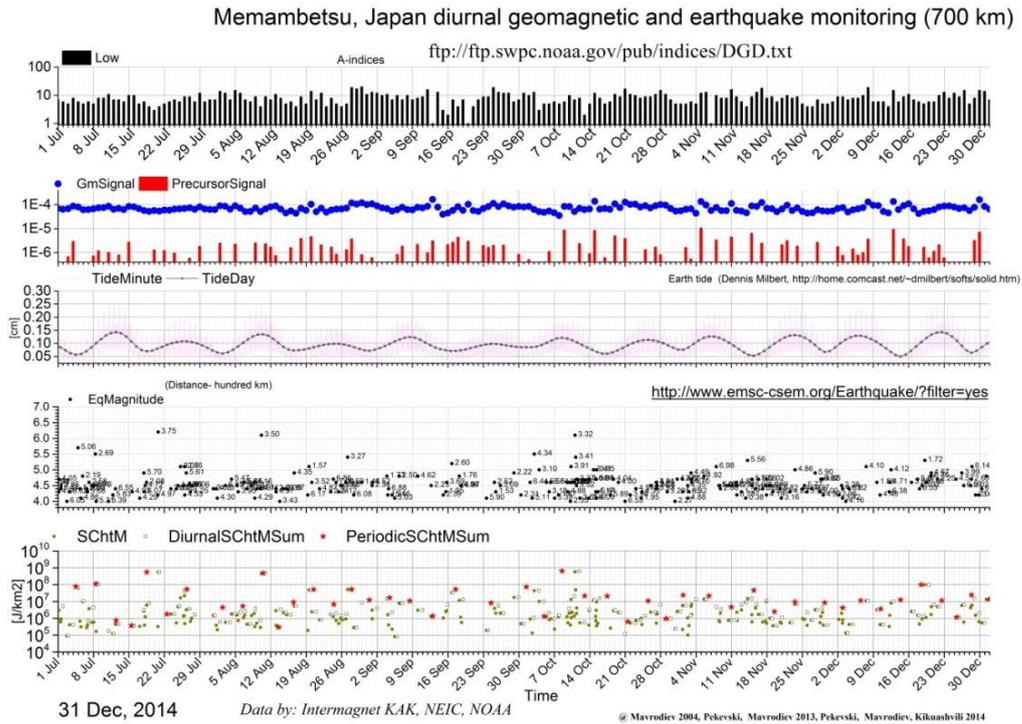

Fig.3. Memambetsu diurnal geomagnetic and earthquakes monitoring for the period Jul 1, 2014 –Jan 1, 2015

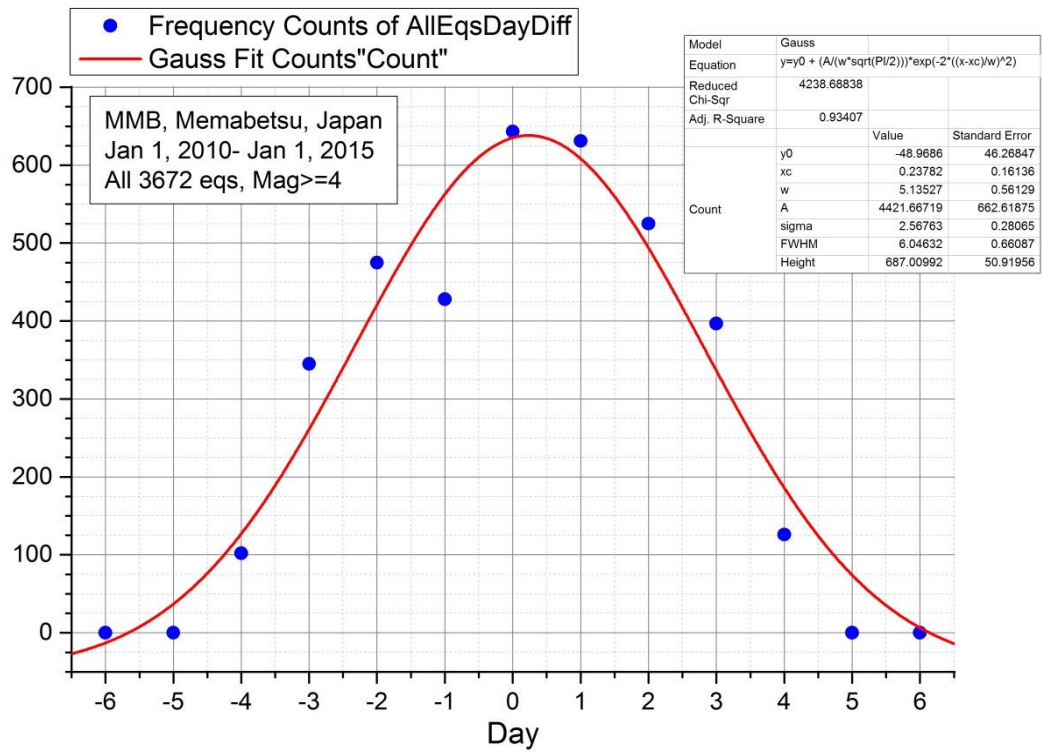

Fig.4. The distribution and its Gauss fit of *DayDiff* for all earthquakes occurred in Memambetsu (700 km) region.

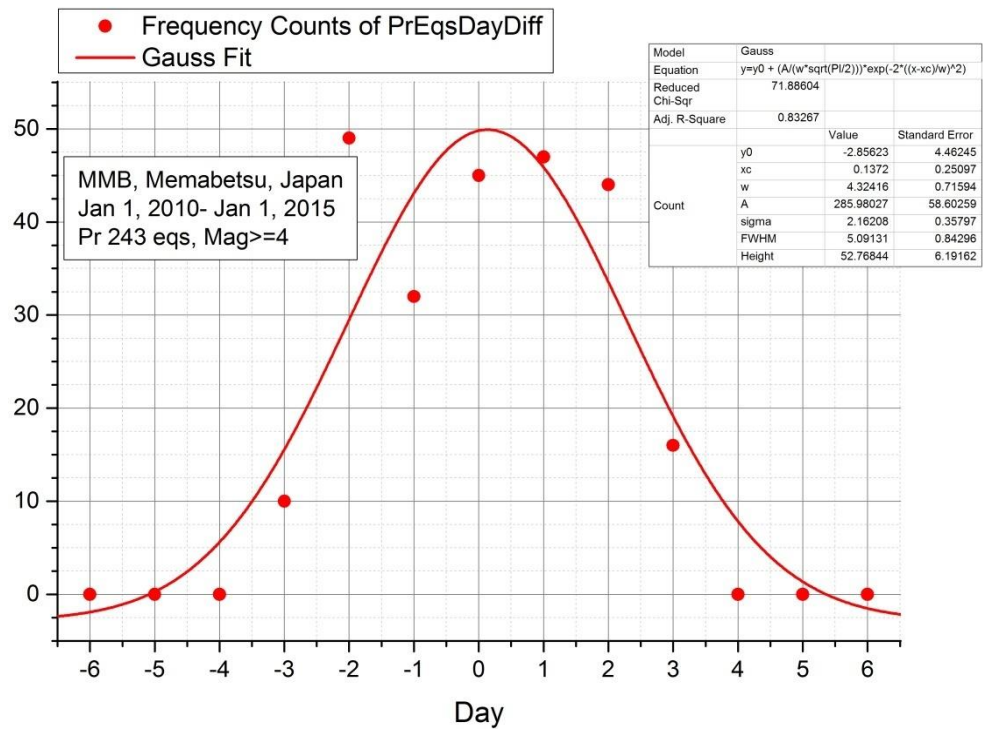

Fig.5. The distribution and its Gauss fit of *DayDiff* for predicted earthquakes in Memambetsu (700 km) region.

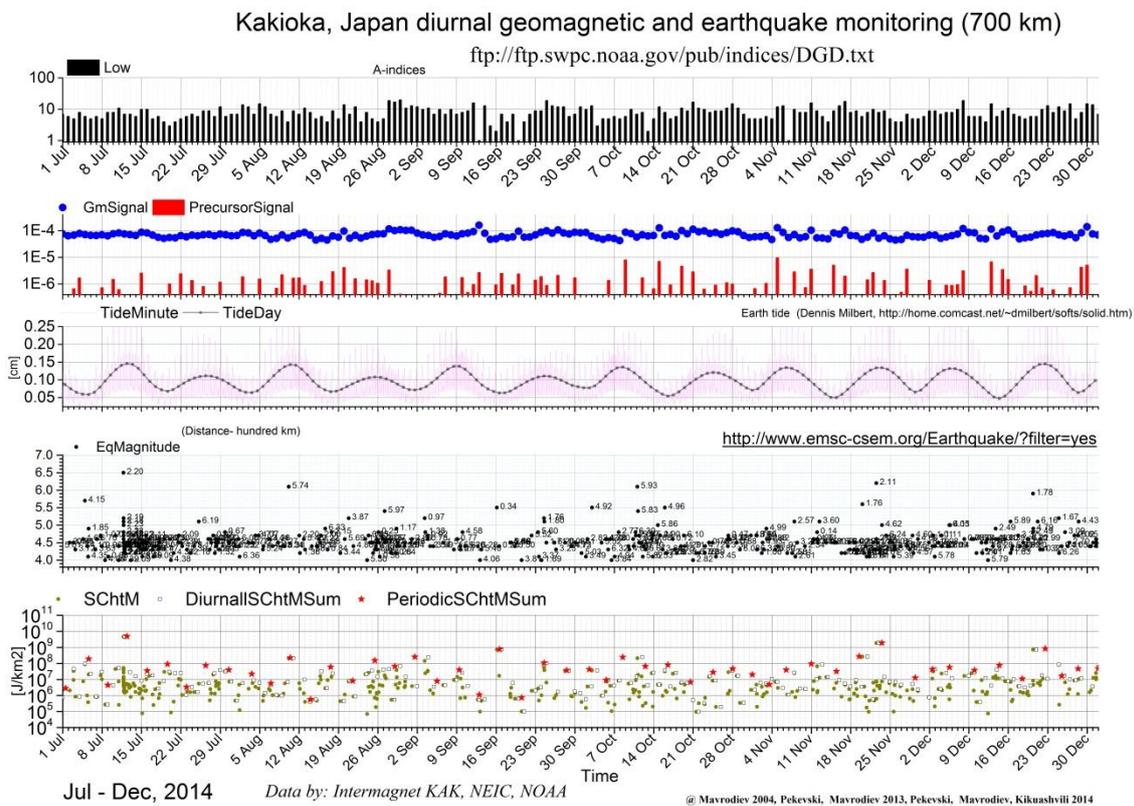

Fig.6. Kakioka diurnal geomagnetic and earthquakes monitoring for the period Jul 1, 2014 – Jan 1, 2015

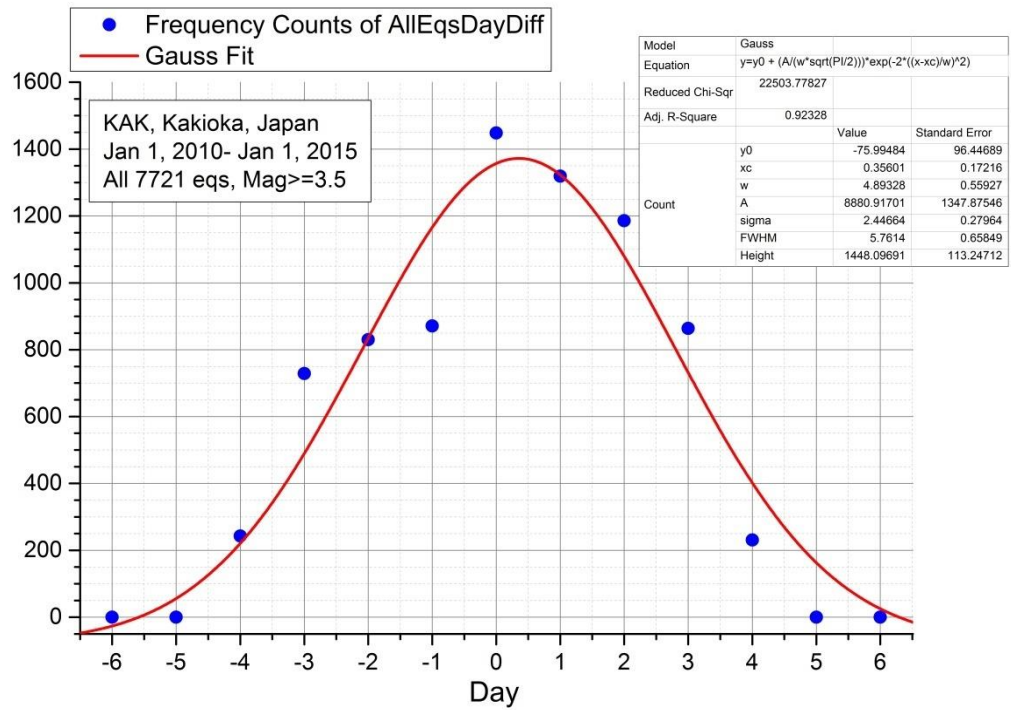

Fig.7. The distribution and its Gauss fit of *DayDiff* for all earthquakes occurred in Kakioka (700 km) region.

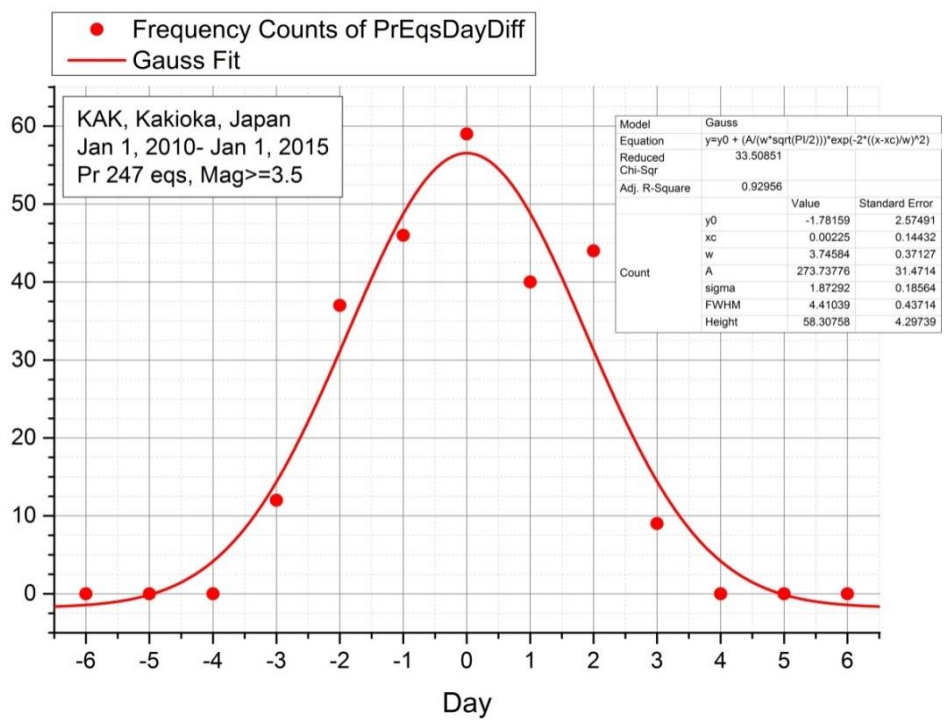

Fig.8. The distribution and its Gauss fit of *DayDiff* for predicted earthquakes in Kakioka (700 km) region.

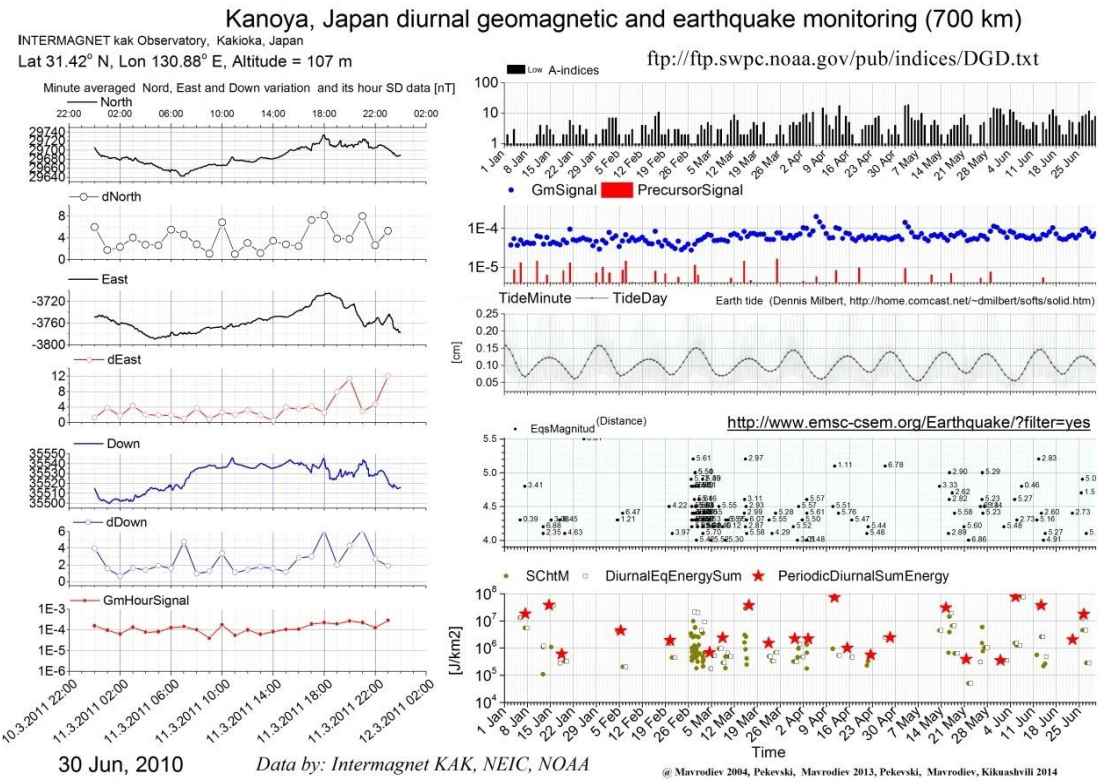

Fig.9.Kanoya diurnal geomagnetic and earthquakes monitoring with geomagnetic field on Jun 30, 2010.

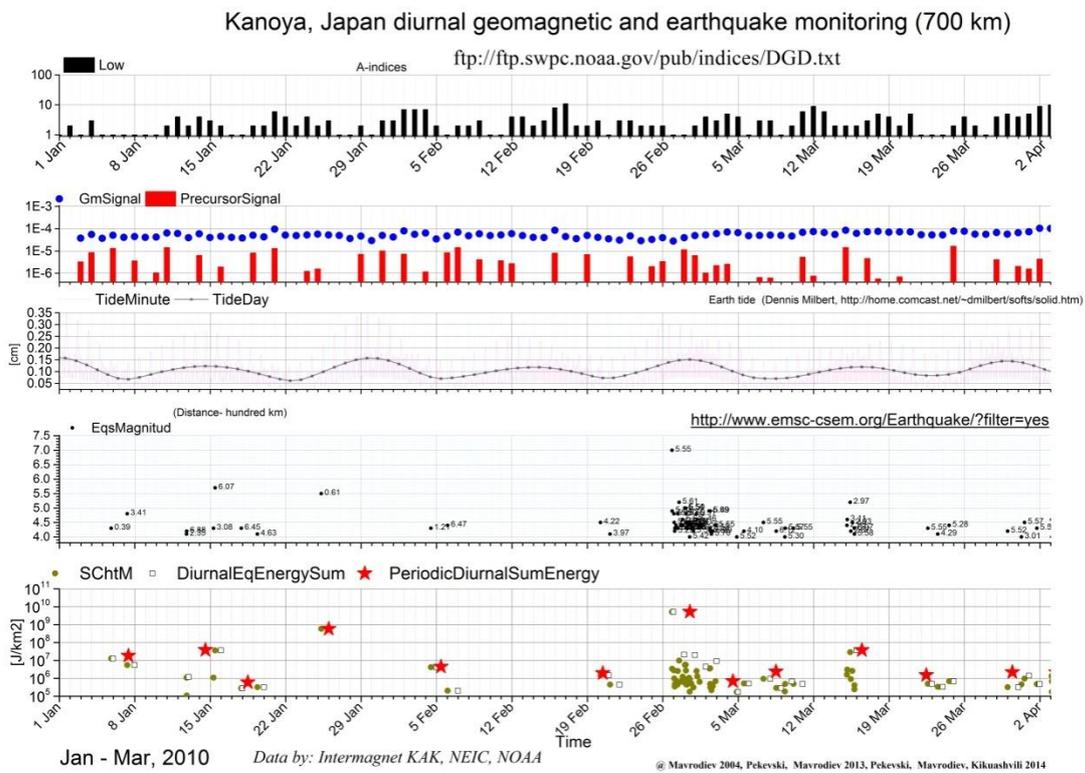

Fig.10.Kanoya diurnal geomagnetic and earthquakes monitoring with geomagnetic field in the period Jan- Mar, 2010.

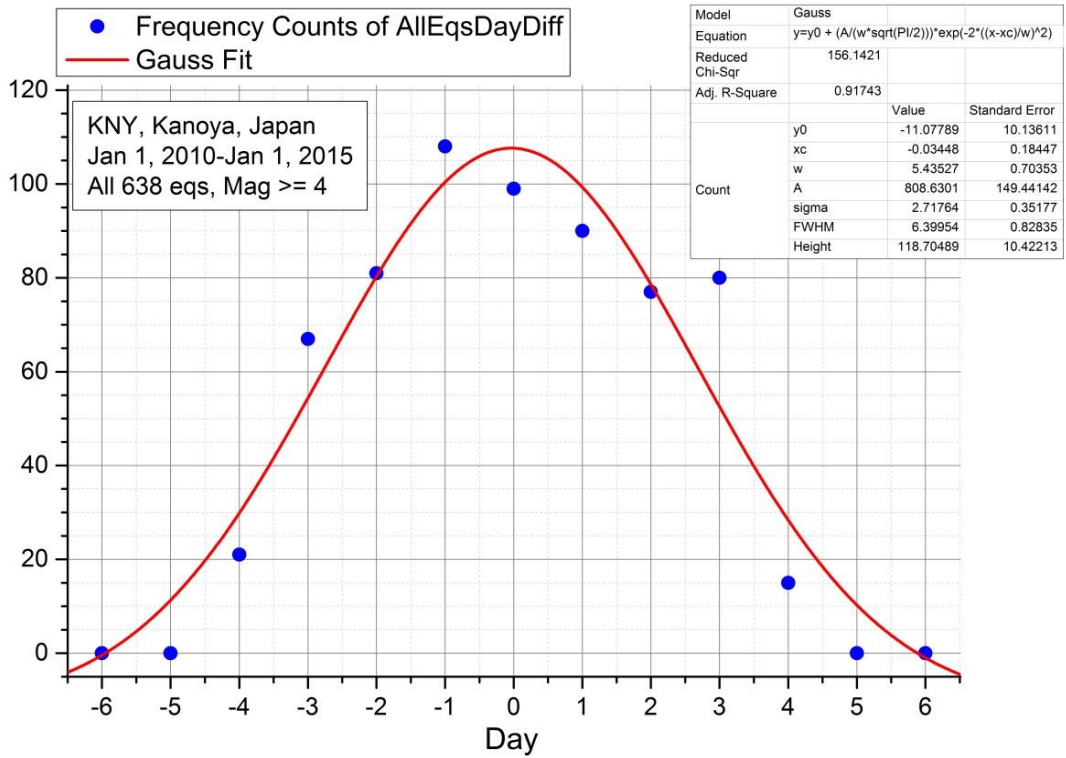

Fig.11. The distribution and its Gauss fit of *DayDiff* for all earthquakes occurred in Kanoya (700 km) region.

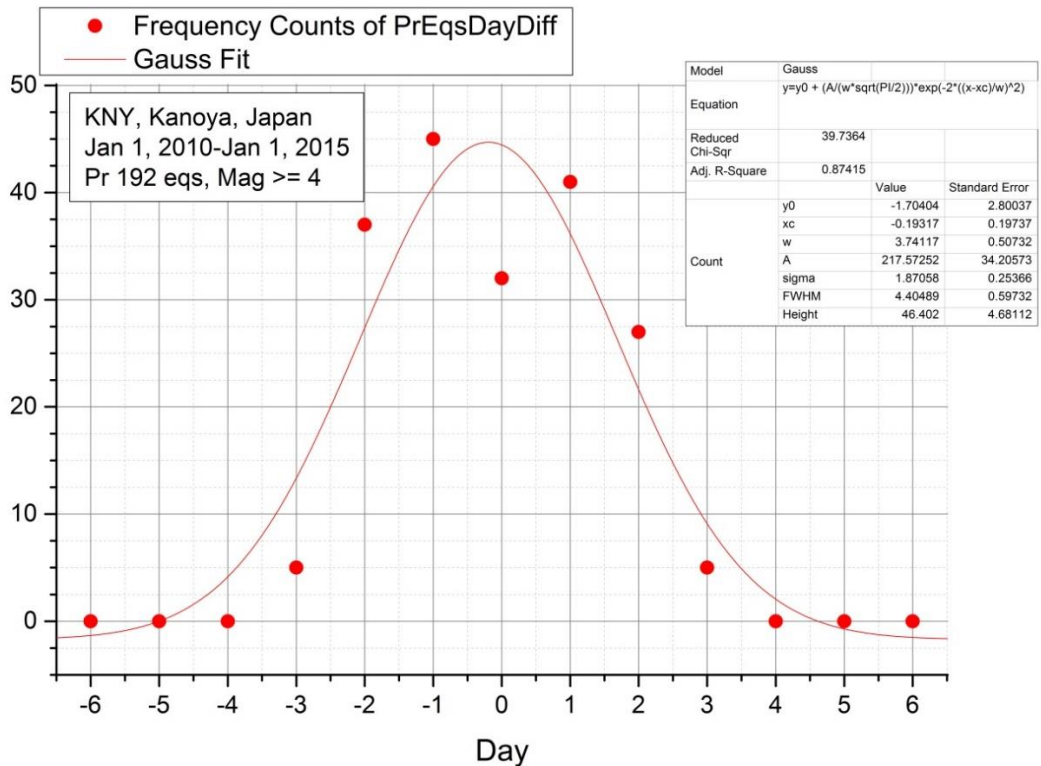

Fig.12. The distribution and its Gauss fit of *DayDiff* for the earthquakes predicted in Kanoya (700 km) region.

In the following **Table1** we present the sum of the variable $S_{ChtM}$ for all earthquakes occurred in the station's region (700 km) and the sum of $S_{ChtM}$ for predicted one, their division in persent and the widths of ***DayDiff*** distribution Gauss fit for all earhquakes, including the ones predicted.

The values of divisions near to 100 % for all the three stations confirm the reliability of the imminent regional seismic activity forecasting. The values of Gauss fit widths can be interpreted as a confirmation of our hypothesis about "predicted" earthquakes: thestrongerthe earthquake is,thehigheris the probability that after the precursor signal it will occur in the region in the time period (+/- 1.97 days) around the time of the followingTide's extreme.

**Table1**

| Station | PrEqs $S_{ChtM}$ Sum [J/km^2] | AllEqs $S_{ChtM}$ Sum [J/km^2] | Pr/All % | Gauss fit width all [day] | Gauss fit width PrEqs [day] |
|---|---|---|---|---|---|
| MMB | 4.01E+12 | 4.11E+12 | 97.6 | 5.14+/-0.56 | 4.32+/-0.72 |
| KAK | 1.48E+13 | 1.68E+13 | 88.1 | 4.89+/-0.60 | 3.75+/-0.37 |
| KNY | 1.96E+10 | 1.98E+10 | 99.0 | 5.44+/-0. 0 | 3.74+/-0.51 |

## 3. Description of precursor signal as a function of earthquake's magnitude, depth and distance

Upon analysing the data for predicted earthquakes presented in Fig. 5, 8 and 12, it was established that there are sixteen earthquakes which are predicted from the signal in two stations simultaneously- See Table 2 in Application 1. So, we have 32 equations for precursor signals, earthquake's magnitude and depth as well as for the distances between the epicenters of the earthquakes occurred and the monitoring points, in which 16 magnitudes and depths have equal values. In thisway we have enough data to formulate the inverse problem- solving the overdetemined system:

$$\boldsymbol{PrecursorSig_i^{Expt} = Th(Mag_i, Depth_i, R_i, A)} \qquad (8),$$

where i=1,…,32, the distance between the epicenter $x_i, y_i$ and the corresponding monitoring point $x_0, y_0$ is $\boldsymbol{R_i = R(x_i, y_i, x_0, y_0)}$ and $\boldsymbol{A}$ ($a_i$, i=1,…n) is a set of unknown digital parameters which define the behaviour of the explicit form of function $\boldsymbol{Th(Mag_i, Depth_i, R_i, A)}$. The discovery of its explicit form and the values of parameter was performed with program code REGN [43]- [46] and its Fortran version is presented in Application 2. One has to note that to facilitate the solution of the system we normed the values of $\boldsymbol{PrecursorSig_i^{Expt}}$ by $10^6$.

The accuracy of description of theexperimet is presented in the following Fig. 13 by variable $\boldsymbol{Res_i}$:

$\boldsymbol{Res_i = (PrecursorSig_i^{Expt} - Th(Mag_i, Depth_i, R_i, A))/PrecursorSig_i^{Expt}}$,
where i=1,..,32.

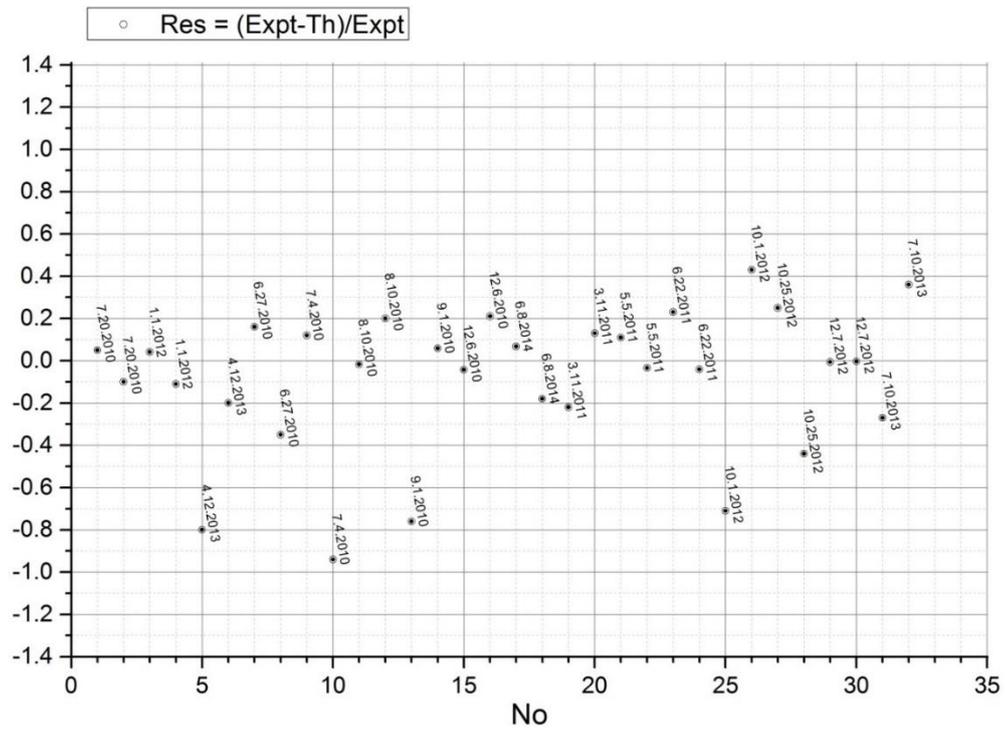

Fig.13. The description of precursor signal as a function of earthquake's magnitude, depth and distance

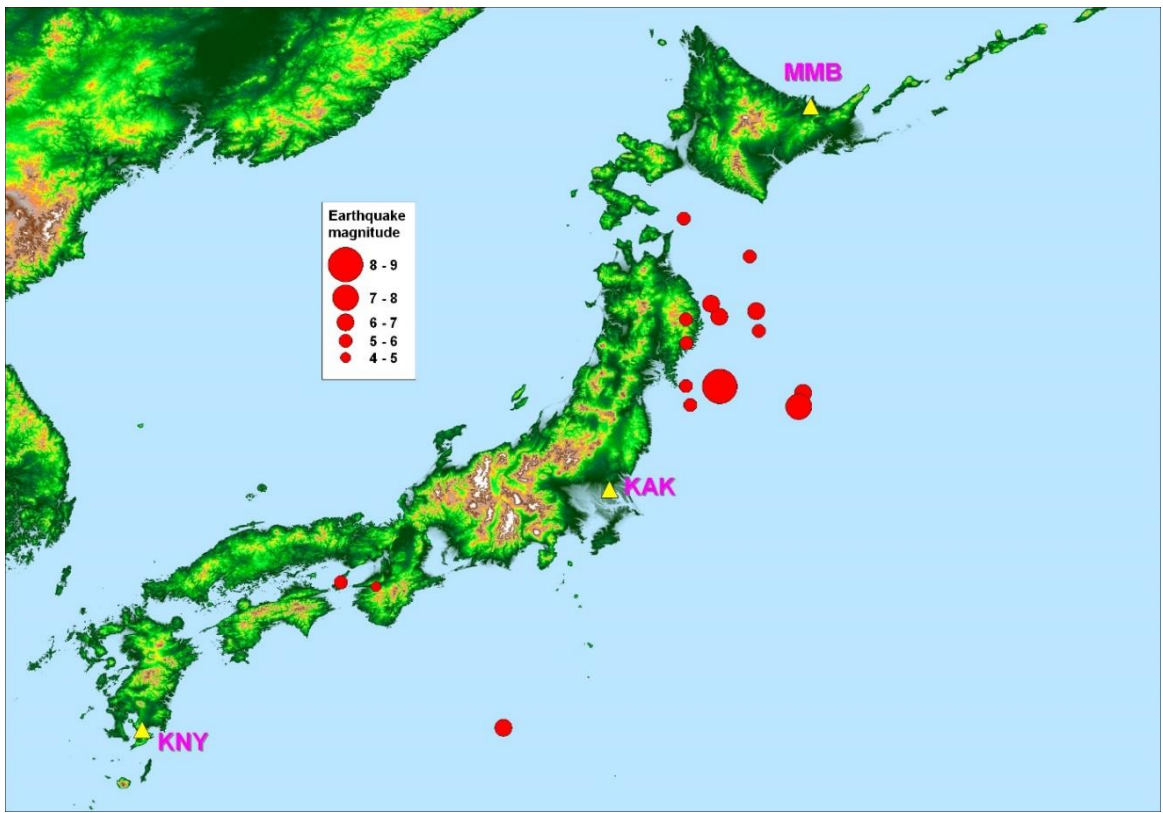

Fig.14. The map illustration of Fig.13.

# 4. Formulation of inverse problem for regional imminent forecasting the magnitude, depth and epicenter coordinates of earthquake

In this section we will present a possibility for solving the inverse problem for the parameters **established** for an incoming earthquake – the time period, magnitude, depth and epicenter coordinate.

If our hypotesis for predicted earthquake is true, it means that after a geomagnetic quake in the following tide extreme with an accuracy equal to +/- 2 days al leas one earthquake in the region will occur.

From the previous section we know the explicit form of precursor signal as function of earthquake magnitude, depth and coordinates of the epicenter- 4 variables.

For calculation the predicted values of this four earthquake's variables we can change the sence of algebraic system (8):

$$PrecursorSig_i^{Expt} = \text{Th}(Mag, Depth, R_i, A) \qquad (9),$$

where i=1,…,*G* - the number of geomagnetic monitoring points.

Now, the solution of the algebraic system **(9)** is the *value of N unkonwn parameters* - the same values of Mag, Depth in every equation and **2.*G*** different values for ***$R_i(x,y, x_i,y_i)$***:

$$N = 2 + 2.G \qquad (10),$$

But the number of equations is G, which means that with a Network for only one precursor it is **not possible to solve** the problem for calculation of Mag, Depth and Coordinates of an incoming earthquake.

The condition to have sufficient data for defining the overdetemined system of equations **(9)** is:

$$2 + 2.G <= P.G \qquad (11),$$

where *P* is the number of precursors (Earth Geomagnetic field, Earth currents field, Borehole water level, Radon concentration, Soil temperature, Atmosphere and Earth core low frequency waves, Ionosphere variability).

**Condition (11) can be satisfied only if *P* >= 3 at *G* > 2.**

**In case** this condition is respected, the first stage of research allows to estimate the epicenter coordinates using simple triangulation, the condition **(11)** is

$$2 + G <= P.G \qquad (12),$$

with solution *P* > 2 and *G* >= 2.

Of course, one has to note that the proposed scheme will take place after the **reliability test of earthquake's precursors** (mentioned in many papers) as Earth electric current, borehole water level, radon concentration, soil temperature, ionosphere behaviour, low frequences wave in the atmosphere and the Earth core will be **performed**.

## Conclusions

The approach proposed for solving the problem of regional imminent "how, where and when" earthquake's prediction does not except the commonly accepted investigations based on seismology, geology, geoelectromagnetism and JPS data.

The reliability test of the Earth currents, Borehole water level, Radon concentration, Atmosphere and the terrestrial low frequency waves as demonstrated in this paper geomagnetic quake reliability for forerecasing the regional seismic activity, after including them in a Regional network, will give data for discovering the explicit forms of different *PrecursorSignal* functions. After collecting enough statistics for a suffucient number of earhquakes occurred in the Network region and solving the overedetermined systems defined from conditions (**9**) we will have data for estimating the **prediction accuracy** for earthquake's time period, magnitude, depth and epifocal coordinates.

## Acknowledgements

I would like to thank my co-authors from BlackSeaHazNet FP7 IRSES (2011- 2014) for cooperation and discutions, Boris Vasilev, JINR, Dubna for the one-component magnetometer, the mathematician and programmer Lubomir Aleksandrov for his code REGN and discussions on the inverse problem methods, Oleg Molchanov and Alexey Sissakyan for many years of fruitful polemics, as well as my late PROFESSORE Vladimir Kadyshevsky, who taught me how to try to graduallyunderstand the physics with analyzing the dynamics and kinemathics of the processes in different time, space and energy scales.

Spacial thanks to Lazo Pekevski, Ludwig Ries, Alexandr Vol and Arie Gilat for the help in the preparation the paper.

## Application 1

Table2

| No | St1St2 | Date MM.DD.YYYY | Lat | Long | Depth km | Mag | SChtM J/km^2 | Distance R 100 km | TimeDiff Day | PrecSignal | Th | Res (Expt-Th)/Expt | Def Expt-Th |
|---|---|---|---|---|---|---|---|---|---|---|---|---|---|
| 1 | KNYKAK | 7.20.2010 | 34.28 | 135.533 | 34.28 | 4.90 | 2.70E+06 | 6.13 | 2.25 | 8.81E+00 | 8.37E+00 | 0.05 | 0.44 |
| 2 | KAKKNY | 7.20.2010 | 34.28 | 135.533 | 34.28 | 4.90 | 4.20E+06 | 4.75 | 2.27 | 4.76E+00 | 5.24E+00 | -0.10 | -0.48 |
| 3 | KNYKAK | 1.1.2012 | 31.456 | 138.072 | 31.46 | 6.80 | 8.20E+08 | 6.96 | 0.18 | 8.00E+00 | 7.67E+00 | 0.04 | 0.33 |
| 4 | KAKKNY | 1.1.2012 | 31.456 | 138.072 | 31.46 | 6.80 | 1.10E+09 | 5.66 | 0.80 | 4.54E+00 | 5.03E+00 | -0.11 | -0.49 |
| 5 | KNYKAK | 4.12.2013 | 34.369 | 134.828 | 34.37 | 5.80 | 8.00E+07 | 5.75 | 2.42 | 2.26E+00 | 4.06E+00 | -0.80 | -1.80 |
| 6 | KAKKNY | 4.12.2013 | 34.369 | 134.828 | 34.37 | 5.80 | 9.30E+07 | 5.29 | 2.42 | 2.77E+00 | 3.34E+00 | -0.20 | -0.57 |
| 7 | MMBKAK | 6.27.2010 | 41.662 | 141.657 | 41.66 | 5.30 | 3.00E+07 | 3.25 | 2.44 | 4.22E+00 | 3.54E+00 | 0.16 | 0.68 |
| 8 | MMBKAK | 6.27.2010 | 41.662 | 141.657 | 41.66 | 5.30 | 3.00E+07 | 3.25 | 2.44 | 2.62E+00 | 3.54E+00 | -0.35 | -0.92 |
| 9 | MMBKAK | 7.4.2010 | 39.697 | 142.369 | 39.70 | 6.30 | 5.70E+08 | 4.93 | 1.41 | 5.50E+00 | 4.84E+00 | 0.12 | 0.66 |
| 10 | KAKMMB | 7.4.2010 | 39.697 | 142.369 | 39.70 | 6.30 | 7.20E+08 | 4.31 | 0.30 | 2.73E+00 | 5.30E+00 | -0.94 | -2.60 |
| 11 | MMBKAK | 8.10.2010 | 39.406 | 143.148 | 39.41 | 5.90 | 1.40E+08 | 5.09 | 0.79 | 4.45E+00 | 4.53E+00 | -0.02 | -0.08 |
| 12 | KAKMMB | 8.10.2010 | 39.406 | 143.148 | 39.41 | 5.90 | 1.80E+08 | 4.39 | 0.19 | 5.83E+00 | 4.68E+00 | 0.20 | 1.20 |
| 13 | MMBKAK | 9.1.2010 | 37.925 | 141.788 | 37.93 | 5.20 | 6.50E+06 | 6.96 | 1.81 | 1.45E+00 | 2.55E+00 | -0.76 | -1.10 |
| 14 | KAKMMB | 9.1.2010 | 37.925 | 141.788 | 37.93 | 5.20 | 3.90E+07 | 2.36 | 1.78 | 4.37E+00 | 4.11E+00 | 0.06 | 0.26 |
| 15 | MMBKAK | 12.6.2010 | 40.904 | 142.967 | 40.90 | 5.70 | 1.30E+08 | 3.49 | 0.79 | 5.67E+00 | 5.91E+00 | -0.04 | -0.24 |
| 16 | KAKMMB | 12.6.2010 | 40.904 | 142.967 | 40.90 | 5.70 | 5.50E+07 | 5.73 | 0.77 | 5.52E+00 | 4.36E+00 | 0.21 | 1.20 |
| 17 | MMBKAK | 6.8.2014 | 39.164 | 141.709 | 39.16 | 5.20 | 8.10E+06 | 5.67 | 2.70 | 4.34E+00 | 4.05E+00 | 0.07 | 0.29 |
| 18 | KAKMMB | 6.8.2014 | 39.164 | 141.709 | 39.16 | 5.20 | 1.70E+07 | 3.53 | 1.62 | 4.70E+00 | 5.53E+00 | -0.18 | -0.83 |
| 19 | MMBKAK | 3.11.2011 | 38.297 | 142.373 | 38.30 | 9.00 | 3.90E+12 | 6.43 | 0.29 | 5.94E+00 | 7.23E+00 | -0.22 | -1.30 |
| 20 | KAKMMB | 3.11.2011 | 38.297 | 142.373 | 38.30 | 9.00 | 1.50E+13 | 3.01 | 0.23 | 9.20E+00 | 8.05E+00 | 0.13 | 1.20 |
| 21 | MMBKAK | 5.5.2011 | 38.17 | 144.032 | 38.17 | 6.00 | 1.30E+08 | 6.39 | 0.18 | 4.20E+00 | 3.74E+00 | 0.11 | 0.46 |
| 22 | KAKMMB | 5.5.2011 | 38.17 | 144.032 | 38.17 | 6.00 | 3.10E+08 | 4.03 | 1.21 | 2.60E+00 | 2.69E+00 | -0.03 | -0.09 |
| 23 | MMBKAK | 6.22.2011 | 39.955 | 142.205 | 39.96 | 6.70 | 2.40E+09 | 4.70 | 0.65 | 4.55E+00 | 3.52E+00 | 0.23 | 1.00 |

| 24 | KAKMMB | 6.22.2011 | 39.955 | 142.205 | 39.96 | 6.70 | 2.60E+09 | 4.50 | 0.69 | 3.49E+00 | 3.64E+00 | -0.04 | -0.14 |
|---|---|---|---|---|---|---|---|---|---|---|---|---|---|
| 25 | MMBKAK | 10.1.2012 | 39.808 | 143.099 | 39.81 | 6.10 | 3.30E+08 | 4.65 | 1.39 | 2.65E+00 | 4.54E+00 | -0.71 | -1.90 |
| 26 | KAKMMB | 10.1.2012 | 39.808 | 143.099 | 39.81 | 6.10 | 3.20E+08 | 4.73 | 2.43 | 7.87E+00 | 4.51E+00 | 0.43 | 3.40 |
| 27 | MMBKAK | 10.25.2012 | 38.306 | 141.699 | 38.31 | 5.60 | 2.80E+07 | 6.58 | 1.98 | 5.73E+00 | 4.30E+00 | 0.25 | 1.40 |
| 28 | KAKMMB | 10.25.2012 | 38.306 | 141.699 | 38.31 | 5.60 | 1.20E+08 | 2.67 | 1.93 | 2.94E+00 | 4.23E+00 | -0.44 | -1.30 |
| 29 | MMBKAK | 12.7.2012 | 37.89 | 143.949 | 37.89 | 7.30 | 1.00E+10 | 6.70 | 0.94 | 5.38E+00 | 5.41E+00 | -0.01 | -0.03 |
| 30 | KAKMMB | 12.7.2012 | 37.89 | 143.949 | 37.89 | 7.30 | 2.70E+10 | 3.82 | 0.91 | 4.09E+00 | 4.10E+00 | 0.00 | -0.01 |
| 31 | MMBKAK | 7.10.2013 | 39.638 | 141.705 | 39.64 | 5.30 | 1.40E+07 | 5.18 | 1.78 | 2.01E+00 | 2.55E+00 | -0.27 | -0.54 |
| 32 | KAKMMB | 7.10.2013 | 39.638 | 141.705 | 39.64 | 5.30 | 2.00E+07 | 4.02 | 1.80 | 4.83E+00 | 3.08E+00 | 0.36 | 1.80 |

Application 2  The FORTRAN version of Precursor signal function.

```
Function PrecSigTh(aMag,Depth,Distance)
IMPLICIT DOUBLE PRECISION(A-H,O-Z)
DIMENSION A(16)
DATA A /0.653118375493643180E+04,  0.239327649144353849E+02, -0.441055930229294688E+03, -0.190062527379474363E+04, &
     -0.195894833103010524E+04, -0.514929656067517226E+04, -0.745560820661331309E+04,  0.421788002467532533E+04, &
      0.420599862430744270E+04,  0.319880390225624069E+04, -0.583971362592100718E+01,  0.536940127973910677E+02, &
      0.510487668346017074E+03,  0.287881656908347106E+00, -0.264988287827522662E+01, -0.614005144491253532E+02/
DepL = dlog(Depth); DisL = dlog(Distance)
Str1 = a(2)*aMag + a(3)*DepL + a(4)*DisL
Str2 = a(5)/aMag + a(6)/(DepL+1.d0) + a(7)/(DisL+1.d0)
Str3 = a(8)/aMag**2 + a(9)/(DepL+1.d0)**2 + a(10)/(DisL+1.d0)**2
Str4 = a(11)*aMag**2 + a(12)*DepL**2 + a(13)*DisL**2
 Str5 = a(14)*aMag**3 + a(15)*DepL**3 + a(16)*DisL**3
PrecSigTh = ( eexp( a(1) + Str1 + Str2 + Str3 + Str4 +Str5 ) )
RETURN
END
```

# References

[1] Main, I. (1999a) Is the reliable prediction of individual earthquakes a realistic scientificgoal?, Debate in *Nature*, http://www.nature.com/ nature/ /earthquake/quake contents.html

[2] Main, I. (1999b) Earthquake prediction: concluding remarks. *Nature debates*, Week 7.

[3] Ludwin, R.S. (2001) Earthquake Prediction. *Washington Geology*, **28**, 3,27, debates.

[4] Pakiser, L. and Shedlock, K.M. (1995) Predicting earthquakes. *USGS*, http://earthquake.usgs.gov/hazards/prediction.html

[5] Geller, R.J., Jackson, D.D., Kagan, Y.Y., Mulargia, F. (1997) Earthquakes cannot be predicted. *Science*, http://scec.ess.ucla.edu/\%7Eykagan/perspective.html

[6] Tamrazian G.P. (1967) Tide-forming Forces and Earthquakes. *Inter. Journal of the Solar System* (IKARUS), **7**, 1, 59-65.

[7] Knopoff, L. (1964) Earth tides as a triggering mechanism for earthquakes. *Bull. Seism. Soc. Am.*, **54**, 1865–1870.

[8] Ryabl, A., Van Wormer, J.D. and Jones, A.E., (1968), Triggering of micro earth-quakes by earth tides and other features of the Truckee, California, Earthquakes sequence of September 1966. *Bull. Seism. Soc. Am.*, **58**, 215–248.

[9] Shlien, S. (1972) Earthquake – tide correlation. *Geophys. J. R. Astr. Soc.*, **28**, 27–34.

[10] Molher A.S. (1980) Earthquake/Earth tide correlation and other features of the Susanville, California, earthquake sequence of June–July 1976, *Bull. Seism. Soc. Am.*, **70**, 1583–1593.

[11] Sounau, M., Sounau, A. and Gagnepain, J. (1982) Modeling and detecting interaction between earth tides and earthquakes with application to an aftershock sequence in the Pyrenees. *Bull. Seism. Soc. Am.*, **72**, 165–180.

[12] Shirley, J. (1988) Lunar and Solar periodicities of large earthquakes: Southern California and the Alaska Aleutian Islands seismic region. *Geophys. J*., **92**, 403–420.

[13] Bragin, Y.A., Bragin, O.A. and Bragin, V.Y. (1999) Reliability of Forecast and Lunar Hypothesis of Earthquakes, Report at *XXII General Assembly of the International Union of Geodesy and Geophysics* (IUGG), Birmingham, UK, 18–30 July.

[14] S. A. Pulinets, D. P. Ouzounov, A. V. Karelin, and D. V. Davidenko, (1915), Physical Bases of the Generation of ShortTerm Earthquake Precursors: A Complex Model ofIonizationInduced Geophysical Processes in the Lithosphere–Atmosphere–Ionosphere–Magnetosphere System, Geomagnetizm i Aeronomiya, 2015, Vol. 55, No. 4, pp. 540–558.

[15] Contadakis, M., Biagi, P. and Zschau, J. (2002) Seismic hazard evaluation, precursory phenomena and reliability of prediction. 27-th EGS General Assembly, Nice, France, http://www.copernicus.org/EGS/EGS.html.

[16] Papadopoulos, G. (2003) 1st International Workshop on Earthquake Prediction Sub commission on Earthquake Prediction Studies (SCE) of the European Seismological Commission scheduled, Athens, Greece, November 2003, http://www.gein.noa.gr/ services/Workshop.htm.

[17] Zhonghao, S.,(1999), Earthquake Clouds and Short Term Prediction, http://quake.exit.com/.

[18] Genrih S.Vartanyan, (2014), Fastdeveloping deformational cycles in the lithosphere and catastrophic earthquakes:Was it possible to prevent the Fukushima tragedy?, Geophysical processes and biosphere, 2014,v. 13 № 2p. 28-63.

[19] Aleksandr Sborshchikov, Genadii Kobzev,Strachimir Cht. Mavrodiev, Georgi Melikadze, (2013), Boreholes water level and earthquake prediction, BlackSeaHazNet series, Vol. 3,p.39,ISBN 978-954-9820-15-7, Sofia.

[20] Dipak Ghosh, Argha Deb, Rosalima Sengupta, (2009), Anomalous radon emission as precursor of earthquake, Journal of Applied Geophysics**,** Volume 69, Issue 2, October 2009, P. 67–81.

[21] A.Gregoric, B.Zmazek, I. Kobal, J. Vapotic, (2011), Radon as earthquake precursor, BlackSeaHazNet series, Vol. 2, p. 175, ISSN 2233-3681, Tbilisi.

**[22]**Varotsos, P. and Alexopoulos, K. (1984a) **,**Physical properties of the variations in the electric field of the earth preceding earthquakes, I**,***Tectonophysics*, 110, 93–98.

[23] Varotsos, P. and Alexopoulos, K. (1984b) Physical properties of the variations of the electric field of the Earth preceding earthquakes. II. *Tectonophysics*, **110**, 99–125.

[24] Constantinos Thanassoulas,(1991) Determination of the epicentre area of the Earthquakes (Ms>6R) in Greece, based on electrotelluric currents recorded by the VAN network. *Acta Geophysica Polonica*, XXXIX, **4**, 373–387.

[25] Varotsos, P., Lazaridou, M., Eftaxias, K., Antonopoulos, G., Makris, J. and Kopanas, J. (1996) Short-term Earthquake Prediction in Greece by Seismic Electric Signals. In: A Critical Review of VAN: Earthquake prediction from Seismic Electric Signals, ed. Ligthhill, Sir J., World Scientific Publishing Co., Singapore, 29–76.

[26] Hayakawa, M. and Fujinawa, Y. (Eds.) (1994) Electromagnetic Phenomena Related to Earthquake Prediction, Terrapub, Tokyo, 677.

[27] Hayakawa, M., Fujinawa, Y., Evison, F.F., Shapiro, V.A., Varotsos, P., Fraser-Smith, A.C., Molchanov, O.A., Pokhotelov, O.A., Enomoto, Y. and Schloessin, H.H. (1994) What is future direction of investigation on electromagnetic phenomena related to earthquakes prediction? In: Electromagnetic phenomena related to earthquake prediction, ed. M. Hayakawa, M. Fujinawa, Y., Terrapub, Tokyo, 667–677.

[28] Hayakawa, M. (Ed.) (1999) Atmospheric and Ionospheric Electromagnetic Phenomena Associated with Earthquakes. Terrapub, Tokyo, 996.

[29] Hayakawa, M., Ito, T., Smirnova, N. (1999) Fractal analysis of ULF geomagnetic data associated with the Guam earthquake on 8 August 1993, Geophys, Res. Lett., 26, 2797–2800.

[30] Strachimir Cht. Mavrodiev and Thanassoulas, C. (2001), Possible correlation between electromagnetic earth fields and future earthquakes. INRNE-BAS, *Seminar proceedings*, 23–27 July 2001, Sofia, Bulgaria, ISBN 954-9820-05-X, http://arXiv.org/abs/physics/0110012

[31] Hayakawa,M. and Molchanov, O. (Eds.) (2002) Seismo Electromagnetics Lithosphere-Atmosphere-Ionosphere coupling**,** Terrapub, Tokyo, 477.

[32] Oike, K. and Yamada, T. (1994) Relationship between shallow earthquakes and electromagnetic noises in the LF and VLF ranges. In: Electromagnetic phenomena related to earthquake prediction, edited by Hayakawa, M. and Fujinawa, Y., Terrapub, Tokyo, 115–130.

[33] Saraev, A.K., Pertel, M.I. and Malkin, Z.M. (2002) Correction of the electromagnetic monitoring data for tidal variations of apparent resistivity. *J. Appl. Geophys*., **49**, 91–100.

[34] Eftaxias, K., Kapiris, P., Polygiannakis, J., Bogris, N., Kopanas, J., Antonopoulos, G., Peratzakis, A., and Hadjicontis, V. (2001) Signature of pending earthquake from electromagnetic anomalies. *Geophys. Res. Lett.*, **28**, 17, 3321–3324.

[35] Eftaxias, K., Kapiris, P., Dologlou, E., Kopanas, J., Bogris, N., Antonopoulos, G., Peratzakis, A. and Hadjicontis, V. (2002), EM Anomalies before the Kozani earthquake: A study of their behavior through laboratory experiments.*Geophys. Res. Lett.*, **29**, 8, 10.1029/2001 GL013786.

[36] Strachimir Cht. Mavrodiev (2004), On the reliability of the geomagnetic quake as a short time earthquake's precursor for the Sofia region. *Natural Hazards and Earth System Sciences*,**4**, p. 433–447. SRef-ID: 1684-9981/NHESS/2004-4-433, © EGU2004.56.

[37] Mavrodiev S. Cht., Pekevski L. and Kikuashvili G. (2013) Results of BlackSeaHazNet Project – FP7 IRSES Project: Extended Executive Summary, Conclusion Workshop, BlackSeaHazNet series, Volume 3, ISSN 2233-3681, ISBN 978-954-9820-15-7, Sofia.

[38] Vladimir G. Kossobokov, Antonella Paresan, G.F. Panza, (2015) On the Operational Earthquake Forecast and Prediction Prblem, SRL,Vol.86, No 2AA, March 2015, p. 287.

[39] NOAA, ftp://ftp.swpc.noaa.gov/pub/indices/DGD.txt.

[40] Dennis Milbert, Earth tide FORTRAN code, http://home.comcast.net/~dmilbert/softs/solid.htm.

[41] Mavrodiev S. Cht. http://theo.inrne.bas.bg/~mavrodi/ (2004)

[42] Mavrodiev, S. , Pekevski, L. , Kikuashvili, G. , Botev, E. , Getsov, P. , Mardirossian, G. , Sotirov, G. and Teodossiev, D. (2015) On the Imminent Regional Seismic Activity Forecasting Using INTERMAGNET and Sun-Moon Tide Code Data. *Open Journal of Earthquake Research*, **4**, 102-113. doi: 10.4236/ojer.2015.43010.

[43] Lubomir Aleksandrov, (1971), Regularized ccomputational pprocess of Newton-Kantorovich type, Comp. Math. and Math., Phys., 11, Vol. 1, 36-43,.

[44] Lubomir Aleksandrov, Strachimir Cht. Mavrodiev, (1976), The Dependence of High-Energy Hadron-Hadron Total Cross-section on Quantum Numbers, Preprint JINR E2-9936, Dubna.

[45] Lubomir Aleksandrov, (1973), The program REGN (Regularized Gauss-Newton iteration method) for solving nonlinear systems of equations, Comm. JINR P5-7259, Dubna,; PSR 165 RSIK ORNL, Oak Ridge, Tennessee, USA,1983.

[46] Lubomir Aleksandrov, Strachimir Cht. Mavrodiev, Alexey N. Sissakian, (2005), On the investigation of some nonlinear problems in high energy particles phenomenology and cosmic ray physics, www.hepi.edu.ge/conferences/